# Single Ion Imaging and Fluorescence Collection with a Parabolic Mirror Trap

Chen-Kuan Chou*, Carolyn Auchter, Jennifer Lilieholm, Kevin Smith, Boris Blinov

*Department of Physics, University of Washington, Seattle, Washington 98195, USA*

(Dated: January 11, 2017)

Single trapped ion qubit is an excellent candidate for quantum computation and information, with additional ability to coherently couple to single photons. Efficient fluorescence collection is the most challenging part in remote entangled ion qubit state generation. To address this issue, we developed an ion trap combining a reflective parabolic surface with trap electrodes. This parabolic trap design covers a solid angle of $2\pi$ steradians, and allows precise ion placement at the focal point of the parabola. We measured approximately 39% fluorescence collection from a single ion with this mirror, and analyzed the mirror optical performance. We observed single ion image spot size of about 3.4 times diffraction limit, improved to 2.8 times diffraction limit with the help of an external deformable mirror. The micromotion of ion is determined to be the limiting factor, and the result is consistent with theoretical calculation.

## I. INTRODUCTION

A quantum bit is very vulnerable to environment perturbations, and entangled state generation, transport, and detection are the most challenging parts in many quantum applications. Entanglement and the means for quantum communication have been demonstrated recently in quantum dots[1, 2], nitrogen vacancy centers in diamond[3, 4], neutral atoms[5], atomic ensembles[6], superconducting qubits[7], and ions[8, 9]. The ion qubit protocol stands out for its long coherence time, and ease of qubit control and detection. The ability to entangle and transport single photons allows the construction of a scalable network for quantum computing[10].

The original radiofrequency quadrupole ion trap designed by Wolfgang Paul is formed by one ring and two end-cap electrodes of hyperbolic shape[11]. Since its invention, there have been various other ion trap designs, including the linear four rods trap[12], fiber tip stylus trap[13], cavity trap[14], semiconductor chip trap[15], and reflective surface trap[16, 17]. To improve optical performance of our spherical mirror trap[16], we built a trap with an optical reflective surface in parabolic shape. When an ion is trapped in the focus of parabola, the mirror is designed to cover half of the $4\pi$ solid angle surrounding the ion, and collimate the reflected photons, which can transport quantum information in free space or through an optical fiber. Compared to the stylus trap combined with a high numerical aperture parabolic mirror[17], our design offers simplicity and robustness of construction and operation.

## II. DESIGN & SIMULATION

The design of the parabolic mirror trap is shown in FIG. 1. The mirror has a focal length of 2 mm, opening of 10.2 mm, and height of 3.2 mm. It has four rectangular slots for laser access. There is a 1.5 mm central through-hole for the needle electrode. The mirror was machined by single point diamond turning by Nu-Tek Precision Optical Corporation, while other parts (including the needle) were machined by University of Washington Physics Instrument Shop. The needle is attached to a linear actuator to allow axial alignment while in vacuum. The mirror is connected to a high voltage radiofrequency potential which, together with a grounded top cross plate and a needle, form the ion trap potential. There are four symmetric stainless steel plates just above the mirror, which are connected to DC bias voltages to allow radial adjustment of ion position. Taking into account the four slots and the hole in the vertex, this mirror covers a solid angle of $2\pi$ steradians. The assembled trap is shown in FIG. 2.

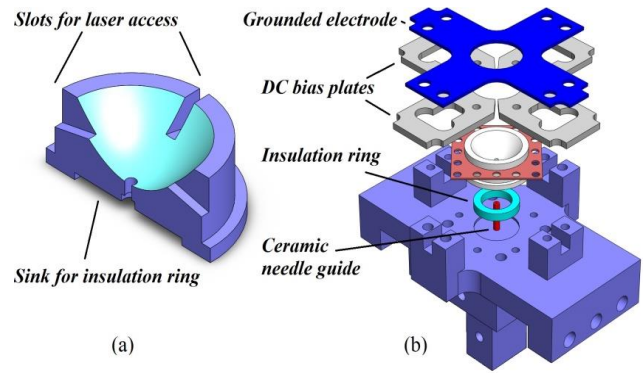

FIG. 1. Mirror (a) and trap assembly (b) design (generated by Solidworks). The mirror was machined by single point diamond turning, other parts (including needle) were machined by UW Physics Instrument Shop.

*hyaline@uw.edu

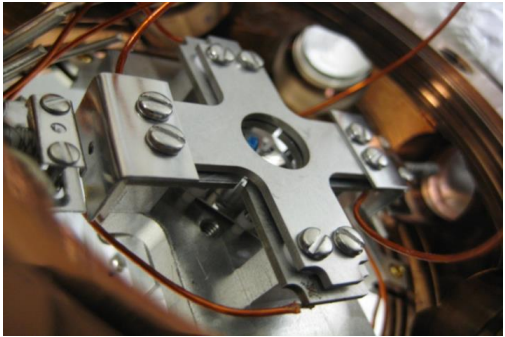

FIG. 2. Assembled parabolic mirror trap placed within vacuum system prior to installing the main viewport. The mirror is in the center, partially obscured in the photo by the ground plate. There are two barium ovens with collimation pin holes in horizontal direction.

We performed a pseudo potential analysis of our parabolic trap using finite element method, and the result is shown in FIG. 3. The origin of the coordinate system is located at the vertex of the parabola. Confinement in the axial direction is tighter than that in the radial direction, and the trapping depth is approximately 0.03 eV for a RF power of 1 watt. The ions are trapped about 0.7 mm above the needle tip, and our simulations show that this distance has little dependence on the RF power and the needle's position along the axis. The trapping potential in the center is almost independent of the optical surface shape, which makes this design universal for different reflector profiles[16].

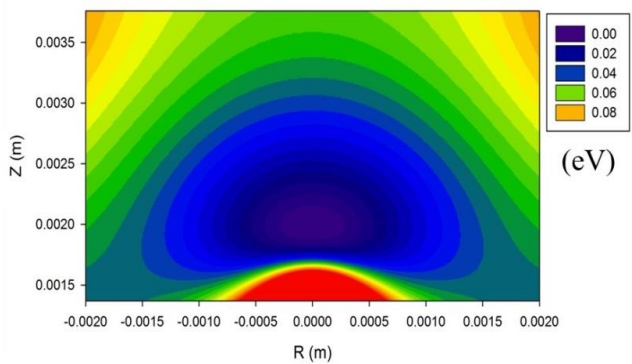

FIG. 3. Pseudo-potential trapping depth simulation. The origin is located at the vertex of the parabola, with Z being the axial coordinate and R the radial. The radial confinement is weaker due to the relatively wide opening of the parabola.

### III. CHARACTERIZATION

To measure the photon collection efficiency of the mirror, we used a single photon counting technique which utilizes 3 energy states of singly ionized $^{138}$Ba[16]. FIG. 4 shows the time sequence of the experiment. The $6P_{1/2}$ excited state has a lifetime of about 8 ns, while the $5D_{3/2}$ meta-stable state has a lifetime of about 82 s, and the $6S_{1/2}$ is the ground state. By switching between the 493 nm and 650 nm laser excitations, we can generate a single 493 nm photon on demand with essentially 100% reliability[16]. The single photons at 650 nm are strongly attenuated by the 495 nm interference filter and result in a negligible background signal.

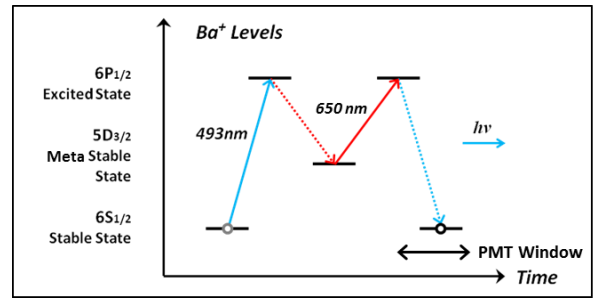

FIG. 4. Single photon generation pattern. The solid lines represent laser excitation and dash lines represent spontaneous emissions. The 493 nm laser is turned on for 500 ns to optically pump the ion into the $5D_{3/2}$ state with essentially 100% efficiency. The 493 nm laser is then switched off, and the 650 nm laser is switched on to generate one and only one 493 nm photon, which is detected with photon counting photo multiplier tube (PMT). The PMT high voltage is gated to further reduce the background.

This pattern can be repeated 1 million times in less than a minute, and we measured 47,675 single photon events per 1 million cycles, i.e. approximately 5% efficiency for producing and detecting single photons, uncorrected for the PMT quantum efficiency and the loss in reflection and transmission. To account for these and other factors in order to measure the photon collection efficiency of the mirror, we measured the dependence of the single photon count rates on the solid angle of collection[16]. We placed a calibrated iris in front of an objective lens (10x Mitutoyo Plan Apo Infinity Corrected long working distance objective) to control the aperture size, and directly (i.e. without using the mirror) focused the ion fluorescence onto a PMT. We used the slope of this calibration curve, shown in FIG. 5, to determine the photon collection efficiency of our mirror to be about 39% of the total $4\pi$ solid angle surrounding the ion. This is a significant improvement over our reflective spherical trap, which collected 25%[16], and custom-made commercial large-NA objective of 10% efficiency[16, 18].

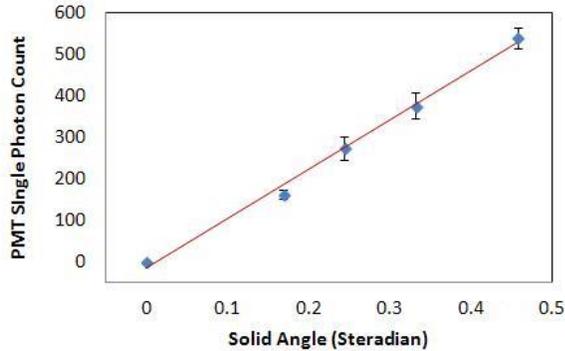

FIG. 5. PMT counts in 1 million cycles for various solid angles set by the calibrated, adjustable aperture. Red line is the linear fit to the data. Error bars are statistical.

The main hoped-for improvement, when compared to our previous spherical mirror design, is the image quality. To examine the imaging quality of our parabolic mirror, we used a spherical lens of 100 cm focal length to focus the collimated fluorescence from a single ion placed in the focus of the mirror. We first moved ion axially using the linear actuator to place it near the focus, then tuned the DC bias voltages on the four plate electrodes to place ion precisely at the focus of parabola. FIG. 6 shows the improvement of the ion image as it is radially shifted.

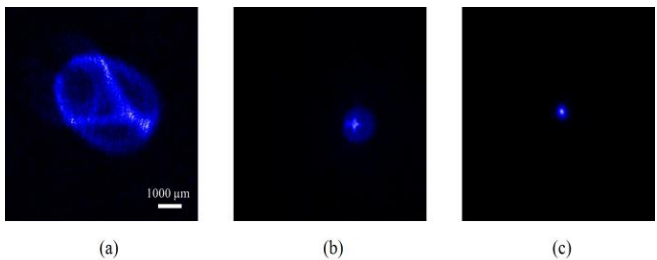

FIG. 6. A series of single ion images as the ion is moved radially using the bias electrodes. The aberration is drastically reduced from (a) as it is moved closer to the focal point of parabola (b and c). The pattern in (a) and (b) comes from the four slots of the parabolic mirror.

The ion image spot size we measured is 1065 µm (defined as $1/e^2$ of peak intensity). Dividing it by the system total magnification of 500, we found 2.1 µm as our optical resolution of a single ion. Comparing to the diffraction-limited Airy disc diameter ($1.22 \times \lambda/NA$), which is ~0.61 µm for our geometry at 493 nm, we determined that the mirror performance is about 3.4 times over the diffraction limit. This was a somewhat disappointing result at first, and we decided determine the reasons.

## IV. ANALYSIS

The ion image spot size is vulnerable to ion defocusing and micromotion, as well as imperfections in the parabolic mirror shape. The first step of our analysis was a ray tracing simulation of point spread function size in various defocusing of ion position. The results are shown in FIG. 7. The accuracy of ion positioning in our experiment is better than 2 µm with the actuator and the bias voltage control. Thus, defocus cannot explain the observed large ion image size. Our next suspect is the imperfection of the parabolic mirror shape. The parabolic mirror is hard to manufacture to precise optical specifications, and measuring its accuracy is difficult as well. Therefore, we adapted a deformable mirror to try to correct the manufacturing imperfection. The deformable mirror we used is a Thorlabs model DMP40. It has 40 independent segments within 11.5 mm pupil diameter, which are arranged as 3 concentric rings. This is ideal for our setup.

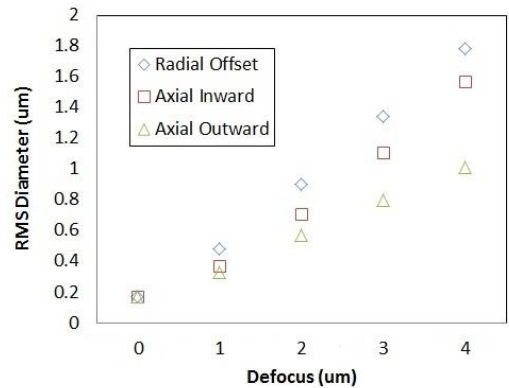

FIG. 7. Point spread function diameter calculated by ray tracing simulation without diffraction. The spot size stays within diffraction limit (~0.61 µm) when ion is out of focus by up to 2 µm axially and 1.5 µm radially.

The imaging optimization is to minimize the ion spot size. With the deformable mirror placed between the parabolic mirror and EMCCD camera, we focused the ion image 400 mm away from deformable mirror, such that the total system magnification is 200. The Thorlabs deformable mirror uses Zernike polynomials to compensate one aberration at a time. The Zernike polynomials are orthogonal polynomials in a unit circle, which simplify the correction process without recursively altering individual segments. The ion image size was about 1.7 µm after optimizing 12 Zernike polynomials parameters for smallest ion spot size. Comparing to the diffraction-limited diameter of 0.61 µm, the overall performance is still about 2.8 times over the diffraction limit. We thus conclude that the image size is not mostly limited by the mirror geometry. The single ion images with and without the deformable mirror (scaled to the same magnification) are shown in FIG. 8. We fitted both ion spot images with Gaussian function. The scale is chosen to make areas under two curves the same, which corresponds to the same photon collection efficiency. The blue curve is a fit to ion image without the deformable mirror, and the red curve is with the deformable mirror. The red curve is clearly sharper with smaller width.

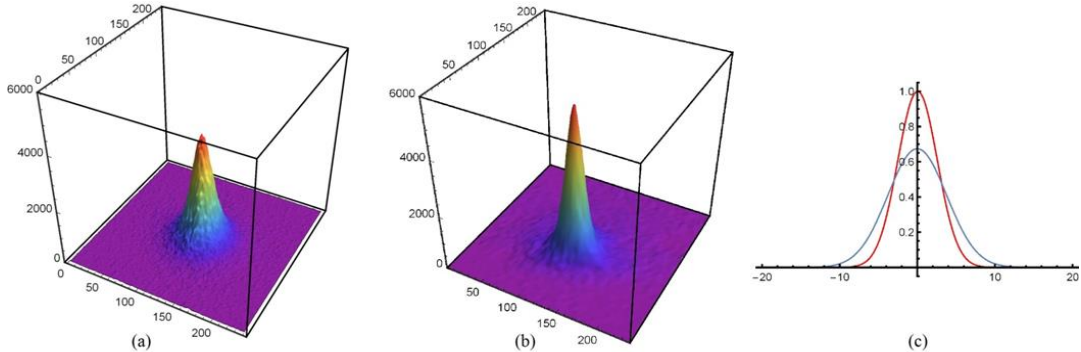

FIG. 8. Ion intensity distributions and their Gaussian fit. (a) is the original image and (b) is the corrected image after deformable mirror optimization. These images are scaled to account for different magnifications, and the ion spot size is improved from 3.4 to 2.8 times over diffraction limit. (c) is the Gaussian function fit of two ion intensity distributions. The blue curve is before-deformable-mirror fit and the red curve is after-deformable-mirror fit. The Gaussian width of blue curve is about 1.35 larger than the width of red curve.

Our ion image spot size is still far from the diffraction limit, and the explanation comes from analyzing the ion micromotion in the trap. The ion's motion in one dimension can be approximated as [12]:

$$x(t) = [x_0 + x_1 \cos(\omega t)][1 + \tfrac{q}{2}\cos(\Omega t)], \qquad (1)$$

where $x_0$ is the ion's static displacement from the trap center due to external electric field, $x_1$ is the ion oscillation amplitude in pseudo potential well, $\omega$ is the secular frequency ($\omega = 2\pi \times 0.3\ MHz$ in our trap in radial direction), $q$ is the dimensionless coefficient related to RF field and trap dimension ($q = 0.1$ in our case), and $\Omega$ is the RF driving frequency ($\Omega = 2\pi \times 12\ MHz$). There are two components in Eq. 1: the slow secular motion with the frequency $\omega$ of the pseudo potential well, and fast driven motion with frequency $\Omega$ of the applied RF field, called the micromotion. When we place the ion in the focus of the parabola radially by applying DC voltages, it is displaced from the center of the RF potential, leading to excess micromotion. The radial displacement is measured to be about 40 μm, which leads to a micromotion amplitude of $x_0 \times q/2 \cong 2\ \mu m$. The secular frequency oscillation amplitude for a Doppler-cooled Ba ion ($m = 137.9\ u$) is $x_1 = \sqrt{\hbar\gamma/m\omega^2} \cong 0.3\ \mu m$, where $\gamma \cong 2\pi * 10\ MHz$ is the natural linewidth of the $6P_{1/2}$ level. We see that, in our setup, $x_1$ is smaller than diffraction limit (0.61 μm), but $x_0$ is quite large, and thus limits our optical resolution. This number is consistent with the 1.7 μm spot size we have measured, which means that the observed the ion image spot size is largely due to the micromotion. It should be pointed out that the micromotion in our case is not one-dimensional. Our DC bias electrodes are four plates which individually cover 1/4 of arc around the mirror opening. This setup creates asymmetrical equipotential lines when we apply DC voltages to more than one electrode. The result is a nonlinear trajectory of ion micromotion, thus our ion spot image is not a clear ellipse along the displacement direction[19].

This was further confirmed by performing measurements of the ion image spot size with different NA of the mirror, by placing a calibrated, adjustable iris after the parabolic mirror reflection. We observed that the ion image spot size did not change noticeably with reduced NA, meaning that our optical resolution is not limited by diffraction or the mirror shape imperfections.

## V. CONCLUSION

In summary, we built a reflective parabolic mirror trap covering 50% of solid angle surrounding an ion trapped at the focus of parabola. The measured photon collection efficiency is 39% measured by single photon counting. We adapted a deformable mirror with 40 independent segments to optimize our ion image quality, and found that the spot size of about 2.8 times over the diffraction limit is due to the ion's micromotion. Our next generation parabolic mirror trap will include in-vacuum piezoelectric actuators to move the needle electrode radially and place the ion in the focus of the parabola while keeping it at the center of the RF pseudopotential, thus minimizing the ion micromotion.

## ACKNOWLEDGEMENTS

The authors wish to thank Thomas Noel and Tomasz Sakrejda, for helpful discussions. This research was supported by National Science Foundation Grant No. 1505326.

## REFERENCES

1. K. D. Greve, L. Yu, P. L. McMahon, J. S. Pelc, C. M. Natarajan, N. Y. Kim, E. Abe, S. Maier, C. Schneider, M. Kamp, S. Höfling, R. H. Hadfield, A. Forchel, M. M. Fejer and Y. Yamamoto, Nature **491**, 421-425 (2012).
2. W. B. Gao, P. Fallahi, E. Togan, J. Miguel-Sanchez and A. Imamoglu, Nature **491**, 426-430 (2012).
3. E. Togan, Y. Chu, A. S. Trifonov, L. Jiang, J. Maze, L. Childress, M. V. G. Dutt, A. S. Sørensen, P. R. Hemmer, A. S. Zibrov and M. D. Lukin, Nature **466**, 730-734.
4. H. Bernien, L. Childress, L. Robledo, M. Markham, D. Twitchen, R. Hanson, Physical Review Letters **108** (4), 043604 (2012).
5. T. Wilk, A. Gaëtan, C. Evellin, J. Wolters, Y. Miroshnychenko, P. Grangier, A. Browaeys, Physical Review Letters **104** (1), 010502 (2010).
6. L. Li, Y. O. Dudin and A. Kuzmich, Nature **498**, 466-469 (2013).
7. C. Eichler, C. Lang, J. M. Fink, J. Govenius, S. Filipp, A. Wallraff, Physical Review Letters **109** (24), 240501 (2012).
8. B. B. Blinov, D. L. Moehring, L.-M. Duan and C. Monroe, Nature **428** (6979), 153-157 (2004).
9. A. Stute, B. Casabone, P. Schindler, T. Monz, P. O. Schmidt, B. Brandstätter, T. E. Northup and R. Blatt, Nature **485**, 482-485 (2012).
10. C. Monroe, R. Raussendorf, A. Ruthven, K. R. Brown, P. Maunz, L.-M. Duan, J. Kim, Physical Review A **89** (2), 022317 (2014).
11. W. Paul and U. B. Physikalisches Institut, Bonn, Germany, Reviews of Modern Physics **62** (3), 531 (1990).
12. D. J. Berkeland, J. D. Miller, J. C. Bergquist, W. M. Itano and D. J. Wineland, http://dx.doi.org/10.1063/1.367318 (1998).
13. R. Maiwald, D. Leibfried, J. Britton, J. C. Bergquist, G. Leuchs and D. J. Wineland, Nature Physics **5** (8), 551-554 (2009).
14. J. D. Sterk, L. Luo, T. A. Manning, P. Maunz, C. Monroe, Physical Review A **85** (6), 062308 (2012).
15. D. Stick, W. K. Hensinger, S. Olmschenk, M. J. Madsen, K. Schwab and C. Monroe, Nature Physics **2** (1), 36-39 (2005).
16. G. Shu, C.-K. Chou, N. Kurz, M. R. Dietrich and B. B. Blinov, JOSA B, Vol. 28, Issue 12, pp. 2865-2870 (2011).
17. R. Maiwald, A. Golla, M. Fischer, M. Bader, S. Heugel, B. Chalopin, M. Sondermann, G. Leuchs, Physical Review A **86** (4), 043431 (2012).
18. D. Hucul, I. V. Inlek, G. Vittorini, C. Crocker, S. Debnath, S. M. Clark and C. Monroe, Nature Physics **11**, 37-42 (2014).
19. T. F. Gloger, P. Kaufmann, D. Kaufmann, M. T. Baig, T. Collath, M. Johanning, C. Wunderlich, Physical Review A **92** (4), 043421 (2015).